\documentstyle[12pt,aaspp4]{article}

\lefthead{Magnetic Dissipation and Electric Currents in the Photosphere}
\righthead{Abramenko}

\begin{document}

\title{Spectrum of Magnetic Dissipation and Horizontal Electric Currents in
the Solar Photosphere}

\author{Valentyna Abramenko}
\affil{Big Bear Solar Observatory, 40386 N. Shore Lane, Big Bear City, CA
92314}

\begin{abstract}

A proxy for horizontal electric currents in the solar photosphere was suggested. 
 For a set of evolving active regions (ARs) observed with {\it Solar and Heliospheric 
Observatory (SOHO)} Michelson Doppler Imager (MDI) in the 
high resolution mode, the dissipation spectrum, $k^2E(k)$, and the spatial 
structure of dissipation, i.e., the Stokes dissipation function 
$\varepsilon(x,y)$, were calculated from  the observed $B_z$ component 
of the magnetic field. These functions allowed us to calculate (a part of) the 
horizontal electric current density in the photosphere.
It was shown that as an active region emerges,  
large-scale horizontal electric currents are gradually generated and
determine a bulk of dissipation. When an active region decays, the 
large-scale horizontal currents decay faster than the small-scale ones.
The density of horizontal currents in active regions is in the range 
of $\langle j_h \rangle \sim (0.008 - 0.028)$ A/m$^2$, that
is compatible with the density of vertical currents in active regions.
We suggest two possible mechanisms for generation of such horizontal currents 
in the photosphere. One of them is the drift motions of charged particles in 
the medium of varying plasma pressure gradient in a horizontal plane at the periphery 
of a sunspot. Such a drift can produce quasi-circular closed horizontal currents around  
sunspots. Another possibility could be an existence of  
horizontal axial current inside a highly twisted horizontal magnetic structure 
laying in the photosphere along the magnetic neutral line.
The horizontal currents may contribute significantly to the dynamics of the 
photosphere/corona coupling, as well as the estimation of non-potentiality of ARs.

\end{abstract}


\section { Introduction}

The ultimate source of energy for the majority of non-stationary processes
in the solar atmosphere is believed to be a dissipation of the free magnetic
energy. This energy, accumulated via plasma motions, manifests as
non-potentiality of the magnetic field or, in other words, as electric currents
that permeate through all levels of the solar atmosphere. Electric currents,
which can be compared to vortex filaments in a turbulent field, tend to be
formed wherever a magnetic field has created a thin layer with both shearing
and stretching. Therefore analysis of the electric current evolution seems to
be a promising approach to better understand processes of generation
and dissipation of the magnetic field as well as accumulation of the free
magnetic energy to be released in flares and CMEs.

In spite of the obvious usefulness of information that electric currents could
provide, it is not trivial to reliably derive them from observed data. 
Rare and mostly state-of-art measurements
of the vector magnetic field in the photosphere can provide the only vertical
component of the current. To derive the horizontal component of the current,
vector magnetic field measurements at several heights in the photosphere are
required, which are not yet available.

Here we present a new approach to indirectly probe horizontal electric currents.
The approach is based on two well known facts, namely, that i) electric current
is a product of the magnetic energy dissipation (Moffatt 1978; Parker
1979; Priest 1982) and ii) the magnetic field in the solar photosphere is in a
turbulent state (Parker 1979, Petrovay \& Szakaly 1993). This allows us to
consider the vertical component of the magnetic field, $B_z$, as a passive
scalar in a turbulent flow and thus to analyze its dissipation in the framework
of the turbulence theory. Spatial and spectral distributions of the dissipation
can then be derived. The former can give us an image of the dissipation over the
active region's area, while the later indicates at which spatial scales the
dissipation dominates. Physical inferences from such consideration can be made
when using the fact that the dissipation of the vertical magnetic field
component is intrinsically related to horizontal electric currents.

In Section 2 of this paper we present two approaches (spatial and
spectral) to derive the dissipation of the $B_z$ component of the magnetic field
and corresponding electric current density. Section 3 is devoted to the 
application of the technique to emerging and decaying active regions.
Discussion is presented in the last section of the paper.

\section{\bf Techniques to derive the magnetic dissipation}

\subsection {Spatial structures of $B_z$ dissipation}

Because the $B_z$ component is transported in the photosphere in the same way as
a passive scalar does (Parker 1979, Petrovay \& Szakaly 1993), one may apply the
Stokes dissipation function (Monin \& Yaglom 1975) to obtain the spatial
structure of dissipation:

\begin{equation}
\varepsilon(x,y) = \frac{1}{2} \left ( \left ( \frac {{\it d}B_z}{{\it d}x}
\right ) ^2 + \left ( \frac {{\it d}B_z}{{\it d}y} \right ) ^2 \right ),
\label{eps}
\end{equation}
where $(x,y)$ is a current point on the horizontal plane and $z$ represents
direction normal to the solar surface.
 In general, the right hand part of Eq.
(\ref{eps}) requires the magnetic diffusion coefficient, $\eta$, which is
unknown. Assuming the quasi-uniform behavior of $\eta$ over the active region
area, we will analyze the spatial distribution of
dissipation as it is derived from Eq. (\ref{eps}) and measured in
$Gs^2/Mm^2$ units.

\begin{figure}[!h] \centerline {\epsfxsize=4.0truein
\epsffile{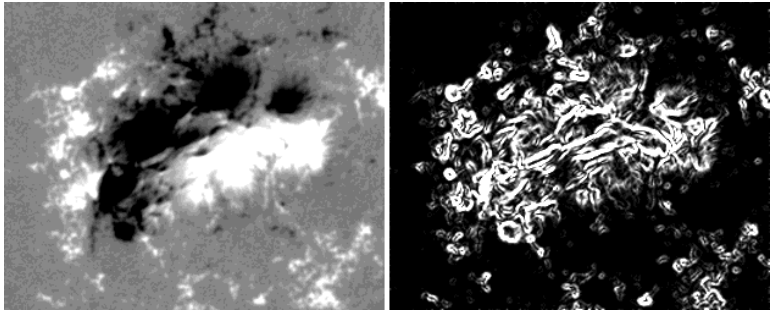}}
\caption{\sf MDI/HR magnetograms ({\it left}) and corresponding dissipation
structures, $\varepsilon(x,y)$, calculated for NOAA AR 0365 observed on 2003
May 26/01:01 UT. North is to the top, west is to the right. The image size is $
156 \times 126$ arcsec. The magnetogram is scaled from -500 G to 500G. The
dissipation map is scaled from 0 to $ 10^5$ G$^2$/Mm$^2$.}
\label{Fig0} 
\end{figure}

A relationship between the dissipation $\varepsilon$ and the squared electric
current, $\bf j^2$, can be easily derived from the general formula for the
squared curl of the magnetic field vector, ${\bf B} \equiv (B_x,B_y,B_z)$:
\begin{eqnarray}
\mu^2 {\bf j}^2 =(\nabla \times {\bf B})^2  = & \nonumber \\
 & \left ( \frac {{\it d}B_z}{{\it d}y} -  \frac {{\it d}B_y}{{\it d}z} \right
) ^2 + \left ( \frac {{\it d}B_x}{{\it d}z} -  \frac {{\it d}B_z}{{\it d}x}
\right )
^2 + \left ( \frac {{\it d}B_y}{{\it d}x} -  \frac {{\it d}B_x}{{\it d}y} \right
) ^2 =   \label{j2} \\
& \varepsilon(x,y)  + \mu^2j_z^2 + f(B_x,B_y,{\it d/dz}). \nonumber
\end{eqnarray}
Thus, dissipation $\varepsilon(x,y)$ represents a part of the horizontal
electric current. The spatial structure of $\varepsilon(x,y)$ is shown in
Figures \ref{Fig0}, \ref{Fig2}, and \ref{Fig3}. Areas of strong dissipation,
and, therefore, strong horizontal currents, tend to be concentrated around
strong magnetic elements. Below we discuss the distribution of horizontal
currents with possible mechanisms for their formation.

Averaging $\varepsilon(x,y)$ over the magnetogram area, we obtain the mean
value of dissipation $\langle \varepsilon \rangle$ associated with $B_z$ component. 

Having the value of $\langle \varepsilon \rangle$ for an active region, we can estimate (a
part of) the horizontal electric current by using Eq. (\ref{j2}):
\begin{equation}
\langle {\bf j}_h^2\rangle =\frac{1}{\mu^2}\langle \varepsilon \rangle.
\label{jtr}
\end{equation}
Here $\mu=\mu_0=4\pi 10^{-7}$ H/m is the vacuum permeability.

\subsection {Spectrum of dissipation of $B_z$}

\begin{figure}[!h] \centerline {\epsfxsize=5.0truein
\epsffile{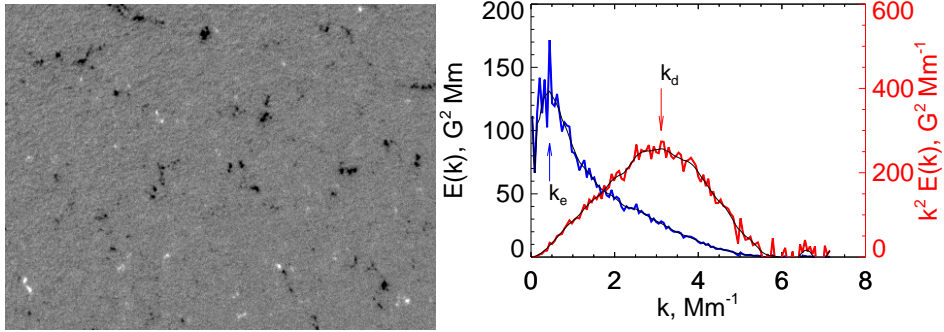}}
\caption{\sf MDI/HR magnetograms ({\it left}) and corresponding energy spectrum,
$E(k)$ ({\it blue curve and right axis}) and dissipation spectrum,
$W(k)=k^2E(k)$ ({\it red curve, left axis}), calculated for a coronal hole area
recorded on 2001 June 5/13:07 UT. Noise influence on the spectra was eliminated.
North is to the top, west is to the right. The image size is $ 266 \times 202$
arcsec. The magnetograms are scaled from -150 G to 150G. The arrows $k_e$  and
$k_d$ mark the maxima of the $E(k)$ and $W(k)$ spectra, respectively.  The
locations of the maxima were derived from 5 point box car averaging ({\it thin
curves}). Peaks of $E(k)$ and $W(k)$ are distinctly separated in the wavenumber
space. }
\label{Fig1} 
\end{figure}

For any measured quantity, which characterizes the
photospheric magnetized plasma, one can calculate the spectrum of energy, i.e.,
a function that shows how the energy (per unit volume) is distributed in the
wavenumber space, $k=2\pi/s$, where $s$ is a spatial scale. The energy spectrum,
$E(k)$ (often referred also as a power spectrum), is essentially a squared
Fourier transform of a measured array (here we consider a one-dimensional
spectrum of a two-dimensional array, see Abramenko (2005) for details of calculation).
For a turbulent medium, the energy spectrum displays a maximum at large scales 
(see Figure \ref{Fig1}, blue curve) followed by a power-law decay at smaller 
scales. The wavenumber where the maximum is reached, marked as $k_e$ ($e$ stands 
for {\it energy}), refers to the scales which bear the maximum energy and where 
the energy input occurs (in our case, this is the largest magnetic elements on 
a magnetogram). 

The spectrum of {\it dissipation} can be derived from the energy
spectrum (Monin \& Yaglom 1975) as 
\begin{equation}
W(k) = k^2E(k). 
\label{W}
\end{equation} 
Usually, for fully developed turbulence at high magnetic Reynolds number (the
ratio of the inertial force to the viscous force), the magnetic dissipation 
becomes significant at high wavenumbers only. As a result, the maximum of the 
dissipation spectrum, $k_d$ (see Figure \ref{Fig1}), is located at higher wavenumbers
than the maximum, $k_e$, of the energy spectrum. The energy interval (which corresponds
to the bulk of energy) and the dissipation interval (which corresponds
to the bulk of dissipation) are separated in the wavenumber space: the
energy is stored mostly at large spatial scales and dissipated at much smaller scales.
The energy cascade is formed from large to small scales, which manifests the
presence of the fully developed turbulent state. Turbulent magnetic diffusion
seems to be a key mechanism for dissipation in this case.

Sometimes the maximum of dissipation, $k_d$, is located close to $k_e$, so that
the energy and dissipation intervals are overlapped. In such a case, the
dissipation regime differs from the one described above. Namely, the bulk of 
dissipation occurs at large scales, and both the ohmic dissipation of large-scale 
currents and the turbulent diffusion at smaller scales contribute into the 
dissipation process. Thus, the dissipation spectrum seems to be useful to diagnose 
the turbulence and the dissipation regime.

Another advantage of the dissipation spectrum arises from its straightforward
relation with the mean squared curl of a field $\bf B$ (Monin \& Yaglom
1975, Biskamp 1993), in other words, with the electric currents: 

\begin{equation}
\langle {(\nabla \times {\bf B} )}^2 \rangle = 2 {\int_0^{\infty}}  k^2 E(k)
dk.
\label{rot}
\end{equation} 
This quantity gives us the mean value of dissipation per unit volume per unit
time. 
We have in our possession the $B_z$ components of the magnetic field only. When
we apply Eq. (\ref{rot}) for $B_z$ by substituting the energy spectrum of $B_z$
as $E(k)$, we acquire the part of dissipation associated with the $B_z$ component,
i.e., the mean value of dissipation per unit area per unit time measured in
units of G$^2$/Mm$^2$. We denoted this value as $\langle W \rangle$:
\begin{equation}
\langle W \rangle = 2{\int_0^{\infty}}  k^2 E(k) dk.
\label{Wsr}
\end{equation} 

One can see that $\langle W \rangle$ and $\langle \varepsilon \rangle$
determine the same physical quantity derived, however, from different
approaches: from spatial and spectral considerations. This circumstance will
allow us to provide two estimations for the mean magnetic energy
dissipation. Besides, the two approaches allow us to study different aspects
of the horizontal electric currents. The current density can be derived as
\begin{equation}
\langle {\bf j}_h^2\rangle =\frac{1}{\mu^2}\langle W \rangle.
\label{jtrW}
\end{equation}

\section{Application to active regions}

We utilized SOHO/MDI (Scherrer at al. 1995) magnetograms recorded in the high
resolution mode. We analyzed areas located at the center of the solar disk (no
farther than 15 degree away from the central meridian), so that the projection effect was
negligible and the direction of $z$ coincides quite well with the line-of-sight
direction.

\begin{figure}[!h] \centerline {\epsfxsize=6.5truein
\epsffile{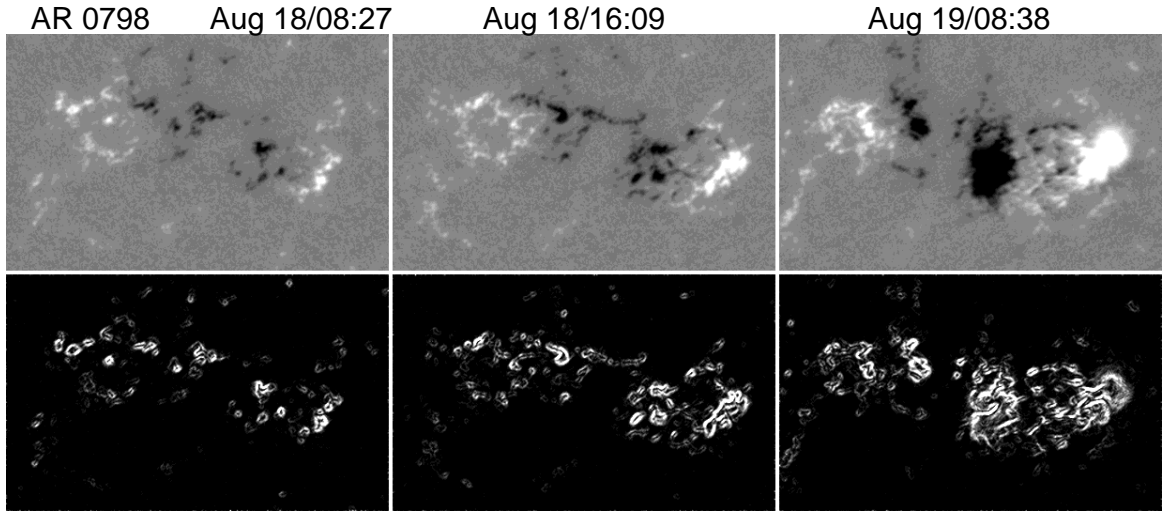}}
\caption{\sf MDI/HR magnetograms ({\it top row}) and corresponding dissipation
structures, $\varepsilon(x,y)$, calculated for emerging NOAA AR 0798 (E8
-W15, S10). North is to the top, west is to the right. The image size is $ 204
\times 126$ arcsec. The magnetograms are scaled from -500 G to 500G. The
dissipation maps are scaled from 0 to $6 \times 10^4$ G$^2$/Mm$^2$.}
\label{Fig2} 
\end{figure}

\begin{figure}[!h] \centerline {\epsfxsize=5.5truein
\epsffile{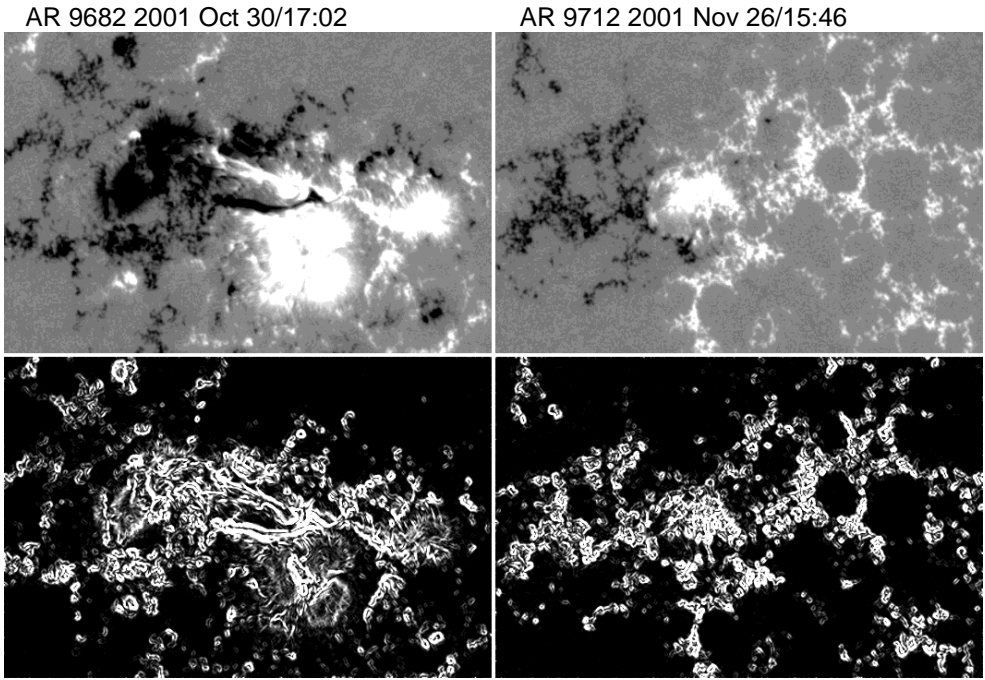}}
\caption{\sf MDI/HR magnetograms ({\it top row}) and corresponding dissipation
structures, $\varepsilon(x,y)$, calculated for a decaying complex NOAA AR 9682/9712. 
North is to the top, west is to the right. The image size is $ 264
\times 174$ arcsec. The magnetograms are scaled from -500 G to 500G. The
dissipation maps are scaled from 0 to $9 \times 10^4$ G$^2$/Mm$^2$.}
\label{Fig3} 
\end{figure}

In this study, we focused on the problem of how the dissipation spectrum and
horizontal currents evolve when an active region emerges and decays. 
An early emergence of an active region can be detected during a 2-day time interval 
when the magnetic structure is in the close vicinity of the central meridian. 
We selected three such cases: NOAA ARs 0314, 9574, 0798 (magnetograms and 
dissipation maps for the last one are shown in Figure \ref{Fig2}). 
However, to detect significant changes at the decay phase, the 
2-day interval is rather short. We decided to analyze the decay of a magnetic structure 
by using observations at two consecutive rotations. For three strong, 
well-developed active regions: NOAA ARs 9682, 9169, 9165, we determined
their remains at the next rotation, namely, NOAA ARs 9712, 9199, 9189, 
respectively. Magnetograms and dissipation maps for the first pair, NOAA ARs 9682/9712, 
are presented in Figure \ref{Fig3}.
As a reference, we also analyzed a coronal hole area shown
in Figure \ref{Fig1}. 

The data for all analyzed magnetograms are compiled in Table 1. 
The first row refers to a coronal hole magnetogram;
three blocks below correspond to three emerging active regions;
three pairs below show the data for decaying complexes. The third (fourth) column, $r_e$ ($r_d$), 
shows the scale where the maximum of the energy (dissipation) spectrum was located, 
the fifth column shows the distance between the dissipation and energy maxima in 
the wavenumber space, ($k_d - k_e$). The next three columns show
the three estimations of the mean value of the magnetic energy dissipation. 
The sixth column, $\langle \varepsilon \rangle$, 
is derived by averaging results from Eq. (\ref{eps}) over a magnetogram. 
The seventh column represents the values $W_n$ calculated from Eq. (\ref{Wsr}) 
when noise was not eliminated in calculation of $E(k)$ (for the noise 
elimination details see Abramenko et al. 2001). 
The eighth column shows the values $W_y$ calculated 
from the same Eq. (\ref{Wsr}), however, noise was preliminary eliminated. Thus, 
for each magnetogram, we obtained values of 
$\langle \varepsilon \rangle$, $W_n$, and $W_y$, which allowed us to use 
Eqs. (\ref{jtr}, \ref{jtrW}) to derive three estimations of the squared horizontal 
electric current density, $\langle {\bf j}_h ^2\rangle$. The average from the 
three estimations $(\langle {\bf j}_h^2\rangle)^{1/2}$ is presented in the last column 
of Table 1. 

\begin{table}[!ht]
\caption{\sf List of studied active regions and calculated parameters}
\footnotesize
\begin{center}
\begin{tabular}{llcrccccc}
\hline
NOAA    & Data            & $r_e$  &$r_d$  &$k_d - k_e$  &$\langle \varepsilon \rangle$&$\langle W_n \rangle$ &$\langle W_y \rangle$   &$(\langle {\bf j}_h^2\rangle)^{1/2}$\\
 AR     &                 &  Mm    &  Mm   &   Mm$^{-1}$ &      G$^2$/Mm$^2$          &    G$^2$/Mm$^2$     &       G$^2$/Mm$^2$    &   Amper/m$^2$            \\
\hline
CH      & 2001 Jun05/13:07 & 14.129 & 2.018 &   2.669     &      3106  	               &       3806           &          1581          &   0.0059 $\pm$ 0.0013    \\
\hline
0314 a  & 2003 Mar13/22:19 & 18.988 & 6.995 &   0.567     &      6929  	               &       6872           &          4500          &   0.0087 $\pm$ 0.0011    \\
0314 b  & 2003 Mar14/11:44 & 44.310 & 10.223&   0.473     &     20471  	               &      19479           &         16878          &   0.0155 $\pm$ 0.0008    \\
0314 c  & 2003 Mar15/11:10 & 44.310 & 14.766&   0.284     &     35149  	               &      33228           &         30396          &   0.0204 $\pm$ 0.0007    \\
\hline			    
9574 a  & 2001 Aug10/05:20 & 31.653 &  4.996&   1.059     &     12726                   &      12708           &         10333          &   0.0122 $\pm$ 0.0007    \\
9574 b  & 2001 Aug10/17:11 & 31.653 & 10.547&   0.397     &     39459  	               &      38421           &         37117          &   0.0220 $\pm$ 0.0003    \\
9574 c  & 2001 Aug11/08:43 & 31.653 & 31.653&   0.000     &     58558  	               &      60528           &         59573          &   0.0275 $\pm$ 0.0003    \\
\hline			    
0798 a  & 2005 Aug18/08:27 & 47.136 &  4.873&   1.156     &      6564  	               &       6570           &          4067          &   0.0085 $\pm$ 0.0011    \\
0798 b  & 2005 Aug18/16:09 & 47.136 &  6.729&   0.800     &      9227  	               &       8833           &          6290          &   0.0101 $\pm$ 0.0010    \\ 
0798 c  & 2005 Aug19/08:38 & 47.136 & 28.264&   0.089     &     13839  	               &      12736           &         10119          &   0.0124 $\pm$ 0.0010    \\
\hline

9682    & 2001 Oct30/17:02 & 63.853 &  7.093&   0.788     &     40802                   &      40082           &         36922          &   0.0223 $\pm$ 0.0006    \\
9712    & 2001 Nov26/15:46 & 63.853 &  2.487&   2.428     &     33839                   &      35281           &         32545          &   0.0207 $\pm$ 0.0004    \\
\hline
9169    & 2000 Sep23/19:00 & 86.307 &  3.547&   1.698     &     59743                   &      59257           &         55694          &   0.0271 $\pm$ 0.0006    \\
9199    & 2000 Oct20/20:53 & 86.307 &  2.785&   2.184     &     25770                   &      26340           &         23946          &   0.0179 $\pm$ 0.0004    \\
\hline			    
9165    & 2000 Sep15/02:00 & 44.848 &  3.281&   1.775     &     44853                   &      43850           &         40661          &   0.0233 $\pm$ 0.0006    \\
9189    & 2000 Oct12/10:03 & 26.908 &  2.862&   1.962     &     15050                   &      15671           &         14051          &   0.0137 $\pm$ 0.0004    \\

\hline
\end{tabular}
\end{center}

\end{table}

\begin{figure}[!h] \centerline {\epsfxsize=5.5truein
\epsffile{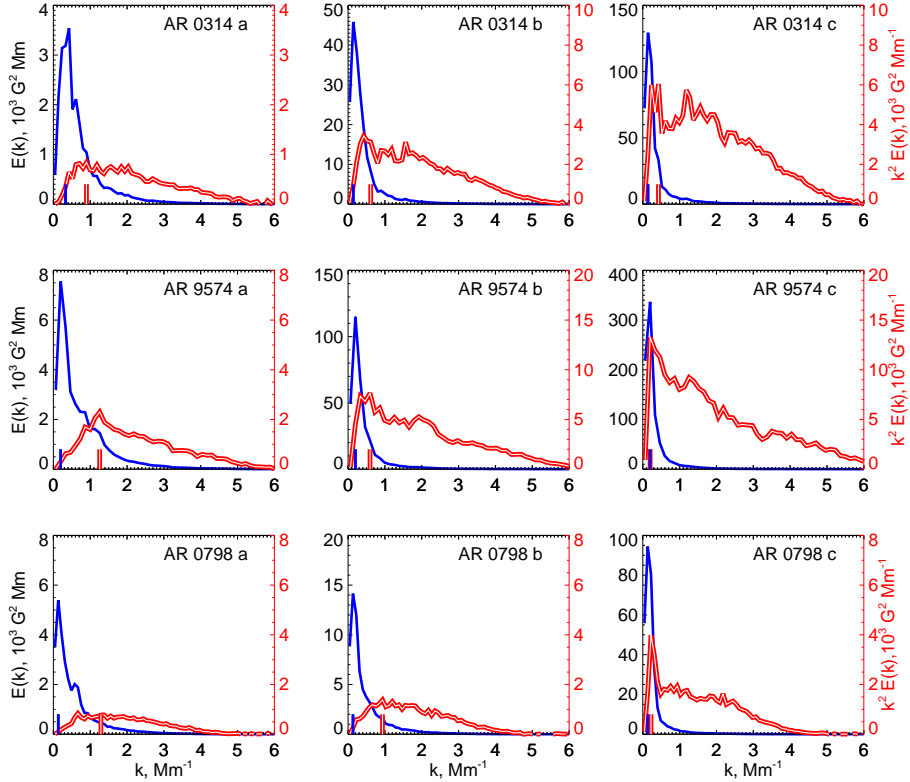}}
\caption{\sf Energy spectra, $E(k)$ ({\it blue lines}), and dissipation spectra,
$k^2E(k)$ ({\it double red lines}), for three emerging active regions. For each AR, 
frames ({\it a, b, c}) correspond to a row with the same mark in Table 1. Vertical 
blue (red) bars mark the maximum of the energy (dissipation) spectrum. Blue bars
 correspond to $k_e$ and red bars correspond to $k_d$. As the active region 
emerges, $k_d$ shifts toward the smaller wavenumbers.}
\label{Fig4} 
\end{figure}

\begin{figure}[!h] \centerline {\epsfxsize=5.5truein
\epsffile{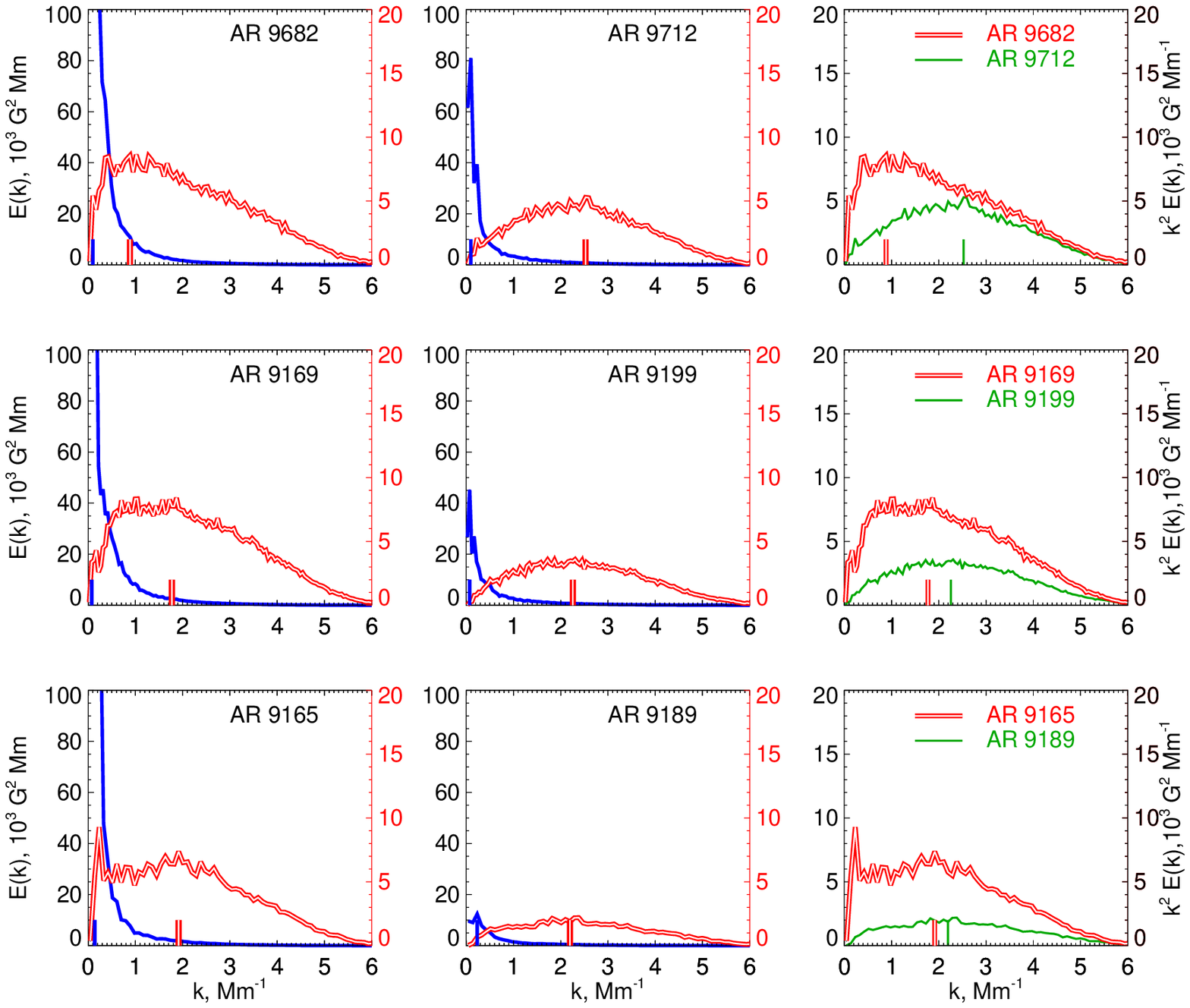}}
\caption{\sf Energy spectra, $E(k)$, and dissipation spectra,
$k^2E(k)$, for three decaying magnetic complexes (three rows). 
The right frame in each row 
shows a superposition of the dissipation spectra at the well-developed 
({\it double red line}) and decaying ({\it solid green line}) state 
of a magnetic complex. As the magnetic complex decays, $k_d$ shifts 
toward the larger wavenumbers.
Other notations are the same as in Figure \ref{Fig4}.}
\label{Fig5} 
\end{figure}

The data from Table 1 show that the coronal hole displays the smallest
energy input scales, $r_e$,  as compared to active regions. This can be 
attributed to smaller typical size of magnetic elements inside coronal holes. 
A non-trivial inference from the CH data is that 
the energy dissipation scale, $r_d$, is smaller and the distance ($k_d - k_e$) 
 is larger than that for active regions. This implies 
that in a CH we observe the state of more developed turbulence as compared to ARs. 
Thus, an assumption of fully developed turbulence seems to be valid for CHs.
The current density is smallest for the CH.

Three emerging ARs show a well pronounced trend as the emergence proceeds. 
Namely, at the very beginning of emergence, we observe the largest distance 
($k_d - k_e$), which resembles the situation in the CH. Later during the emergence 
this distance decreases by means of shifting of $k_d$ toward the small wavenumbers.
The intervals of the energy and dissipation become more and more overlapped 
(see Figure \ref{Fig4}, frames from {\it a} to {\it c}) and the current density 
increases (the last column in Table 1). The observed fact, that during the emergence 
the bulk of dissipation shifts toward larger scales, implies the formation 
of large-scale strong currents. In the map of spatial distribution of dissipation 
(see Figure \ref{Fig2}) we see formation of bright contours (frequently closed) 
surrounding the magnetic flux tubes. Their size increases as the AR emerges.

Three decaying magnetic complexes showed an opposite behavior, see Figure \ref{Fig5}.
As the magnetic structure decays, the maximum of dissipation shifts toward the 
smaller scales, large-scale currents decay, the current density decreases and 
the entire picture begins to resemble what we saw for a CH. Disappearance of large-scale 
dissipation features is distinctively visible in Figure \ref{Fig3} (compare 
left and right frames in the bottom row.)

The data obtained allow us to estimate the upper boundary of the plasma conductivity,
 $\sigma=\tau/l^2$. Here $l$ is a typical scale of decayed current and $\tau$ 
is a typical dissipation time. The data for NOAA ARs 9682/9712 pair show that 
currents of scales about 7 Mm  dissipated during $\tau <$ 27 days. This gives us 
$\sigma < 5\times 10^{-8}$ s/m$^2$, which is four orders of magnitude lower than 
the classical ohmic conductivity for the photosphere.
The magnetic diffusion coefficient in the photospheric plasma of active regions: 
$\eta=l^2/\tau$ (Priest 1982, here $l$ is a typical scale of decayed magnetic element and $\tau$ 
is a typical decay time). For magnetic elements of 7 Mm  dissipated during 
$\tau <$ 27 days we derive $\eta > 2\times 10^7$ m$^2$/s, which is four 
orders of magnitude higher than the classical diffusion coefficient for the photosphere.

\section { Discussion}

The study presented here suggests that in an active region there exist large-scale 
horizontal electric currents, which appear and grow during 1-2 days as the AR
emerges and slowly decay, during a couple of weeks, when the AR decays. 
A bulk of magnetic dissipation of the $B_z$ component is associated with these 
large-scale currents rather than with small-scale currents. 

This inference recalls the publications where the 
photospheric and chromospheric magnetic fields were compared  (Abramenko et al. (1992), 
Choudhary et al. (2002), Leka \& Metcalf (2003), Balasubramaniam et al. (2004);
 see also a resent review by Nagaraju et al. (2008) and references in).
Their results  
suggest that the stronger the line-of-sight magnetic field, the faster it 
decreases with height. Nagaraju et al. (2008) argued that for the photospheric 
field stronger than 700 G, the chromospheric field values "are much weaker 
than what one gets from the linear relationship [between the photospheric and 
chromospheric fields] and also from those expected from the extrapolation 
of the photospheric magnetic field" in the potential approximation. 
Leka \& Metcalf (2003) observed that the potential field of NOAA 8299 extrapolated from the 
photosphere to reasonable heights (lower than 2 Mm) statistically exceeds the observed 
 chromospheric field. 
In 1992 Abramenko and co-authors reported that for NOAA 6280 the chromospheric field 
measured in $H_{\beta}$ spectral line is lower than the extrapolated potential 
field. They also found that the vertical gradient of the line-of-sight magnetic fields 
is higher for strong fields. 
Note, that the above results were obtained from different types of measurements, 
different spectral lines, different instruments, etc., and still they are 
consistent with each other.
Several reasons were suggested to explain the effect (see Nagaraju et al. (2008) for review).
However, only one of them addresses the effect of horizontal electric current formation.
Namely, in Abramenko et al. (1992) the conclusion was made that the possible reason 
for such a systematical deviation between the observed and potential fields
is the presence of horizontal electric currents enclosing sunspots. The magnetic fields of 
such currents are predominantly vertical and directed opposite to the original 
field of a sunspot, so that superposition of the two fields results in the 
fast weakening with height of the observed field.

We suggest that one of the possible mechanisms for formation of such horizontal currents
could be the drift motions of charged particles in the medium of varying plasma 
pressure gradient in a horizontal plane. Presence of such a medium is unavoidable 
at the periphery of a sunspot. Such a drift can produce quasi-circular closed 
horizontal currents around sunspots. This suggestion is in part confirmed by 
the imaging of the horizontal currents, see Figures \ref{Fig0}, \ref{Fig2}, 
\ref{Fig3}: in all of them, quasi-circular closed bright structures around 
sunspots are noticeable. 

Another mechanism for the horizontal currents formation can be suggested. 
Often inside strong active  regions with stressed and sheared neutral line 
one observe strongly elongated magnetic features of both polarities along 
the neutral line, see, for example, Figures \ref{Fig0}, \ref{Fig3}. The most 
obvious way to explain such a phenomenon is to suggest a presence of highly 
twisted horizontal magnetic structure laying in the photosphere. A strong 
electric current along the axis of a magnetic helical tube is unavoidably present. 
A straight long electric current $I$ generates an azimuthal magnetic field 
$B$ at a distance $r$ from the cur rent axis: $B=2I/cr$, where $B$ is in Gauss, 
$I$ in the CGSE system of units, and $c$ is the light velocity. The data for 
NOAA AR 9682 (see Figure \ref{Fig3}) shows that an azimuthal magnetic field 
of about 500 G is generated at a distance of approximately 3 Mm from the axis 
of the horizontal current channel laying along the neutral line. This gives us 
an estimation of the current in the channel as $0.75 \cdot 10^{12}$ A and the 
density of the current as $0.026$ A/m$^2$, which is compatible with the 
magnitude given in Table 1 for this active region.

Emergence of a horizontal helical flux rope along a neutral line was recently 
reported by Okamoto et al. (2008) from vector magnetic field observations with 
the Solar Optical Telescope (SOT) on board the {\it Hinode} satellite.

Taking into account all the above mentioned results, we suggest that strong 
{\it horizontal} electric currents are present in the solar atmosphere, and they 
occupy a large interval of heights: at least, from the photosphere to 
 chromosphere. Their density, $\langle j_h \rangle \sim (0.008 - 0.028) $ A/m$^2$, 
is quit compatible with the density of vertical currents in the photosphere, 
see, e.g., Abramenko et al. (1991): $\langle j_z \rangle \sim 0.06 $ A/m$^2$; 
Wheatland  (2000): (0.009 - 0.015) A/m$^2$. The horizontal currents may contribute 
significantly into the dynamics of the photosphere/corona coupling, as well 
as into estimation of non-potentiality of ARs. They have to be closely connected 
with transverse plasma motions, with creation of shear motions. From this standpoint, 
the recently suggested new approach to study plasma velocities (Nagaraju et al. 2008) 
seems to be very promising.

One more result of the above study deserves to be briefly discussed. 
The shifting of the maximum of the dissipation spectrum as an active region emerges 
and decays indicates that the turbulence regime varies with the evolution of the AR.
Namely, as an active region emerges, the state of fully developed turbulence 
becomes gradually replaced by a state of under-developed turbulence.
The state of under-developed turbulence in mature active regions 
still bears a signature of a turbulent cascade, that is confirmed by 
{\it gradual} decrease of energy along the energy spectrum.
High value of the magnetic diffusion coefficient obtained 
here implies that the turbulent diffusion is significant 
in well-developed active regions and is responsible to the energy cascading. 
However, the presence of a bulk of dissipation at large scales suggests that 
the ohmic dissipation of large-scale currents might compete with the turbulent 
diffusion in formation of a dissipative regime in the photosphere of 
well-developed active regions. As the active region decays, the large-scale 
currents decay faster, as compared to the small-scale currents, and the 
fully-developed turbulence tends to be restored.

Author is thankful to Prof. Lennard Fisk for a suggestion to use the dissipation 
spectrum as a proxy for the mean squared curl, to Drs. Vasyl Yurchyshyn, 
Gregory Fleshman, Haimin Wang for helpful discussions, to Erika Norro and Aaron Coulter 
for their help with language issues. 
This work was supported, in part, by NASA NNX07AT16G grant and 
NSF grant ATM-0716512.

{}

\end{document}